\documentclass{PoS}
\usepackage{booktabs}
\usepackage{url}

\def\arxiv#1{\href{http://arxiv.org/abs/#1}{{\tt arXiv:#1}}}
\def\astroph#1{\href{http://arxiv.org/abs/astro-ph/#1}{{\tt astro-ph:#1}}}

\title{The exceptional TeV flaring activity of the blazar 1ES 1959+650 in 2015 and 2016 as observed with VERITAS}

\ShortTitle{VERITAS observations of 1ES 1959+650 in 2015 and 2016}

\author{\speaker{M.~Santander},$^a$ on behalf of the VERITAS$^{\dagger}$ Collaboration \\
\llap{$^a$} Barnard College, Columbia University, New York, NY, USA \\
E-mail: \email{santander@nevis.columbia.edu} \\
\llap{$^{\dagger}$} http://veritas.sao.arizona.edu \\
}

\abstract{The high-synchrotron-peaked blazar 1ES 1959+650 was among the first extragalactic sources detected in the very high energy gamma ray band (VHE, E > 100 GeV). In October 2015, the source entered an extended period of activity that continued through July 2016, during which several strong VHE flares were observed. This flaring activity in the TeV band was accompanied by a strong increase in the optical, X-ray, and GeV gamma-ray flux of the source, surpassing its brightest recorded flux states. The VERITAS telescope array performed observations of 1ES 1959+650 between October 2015 and June 2016, and detected the source multiple times at a flux higher than the Crab nebula flux in the TeV band, representing the brightest flares of this object since 2002. We here present results from the analysis of 32 hours of VERITAS observations obtained during this period and as well as a contemporaneous multi-wavelength observations in the optical, X-ray, and GeV gamma-ray bands.}

\FullConference{35th International Cosmic Ray Conference --- ICRC2017\\
		10--20 July, 2017\\
		Bexco, Busan, Korea}

\begin{document}

\section{Introduction}

The high-synchrotron-peaked blazar 1ES 1959+650 ($\alpha_{2000}$: 19$^{\mathrm{h}}$ 59$^{\mathrm{m}}$ 59.8$^{\mathrm{s}}$, $\delta_{2000}$: +65$^{\circ}$ 08$^{\mathrm{m}}$ 55$^{\mathrm{s}}$) at a redshift of z=0.047 is a well known gamma-ray source, the seventh object detected in the very-high-energy range (VHE, $E > 100$ GeV) and only the third of the BL Lac type. Observations of the source in a lower state have reported a flux level of $\sim23$\% that of the Crab nebula flux (C.U.) above 1 TeV, or (3.97 $\pm$0.37) $\times 10^{-12}$ photons cm$^{-2}$ s$^{-1}$~\cite{LowState}. Since its first detection in the VHE range~\cite{Detect1, Detect2}, 1ES 1959+650 has exhibited several strong flaring episodes surpassing the flux of the Crab nebula, making it one of only six BL Lacs known to do so. 

The most notable of these flares was detected in 2002 by the Whipple telescope, when a 4--5 C.U. flare was observed without an accompanying flare in X-rays~\cite{Orphan}. So-called ``orphan'' gamma-ray flares represent a challenge for one-zone synchrotron self-Compton (SSC) models of gamma-ray production in blazars and several models were suggested to explain the observations. Some models included a hadronic origin for the gamma-ray flux, where the gamma-rays are due to the decay of neutral pions produced in cosmic-ray interactions. These interpretations were further motivated by the \emph{a posteriori} detection of three neutrino events with the AMANDA telescope in temporal and spatial coincidence with the 1ES 1959+650 flare~\cite{AMANDA}. The blazar has remained an interesting target for neutrino telescopes such as IceCube, and while no significant detection of a neutrino excess has been found, the source location exhibits the largest over-fluctuation of neutrino candidate events in the point-source searches in the northern sky~\cite{IceCube}.

\begin{figure}[!th]
\centering
\includegraphics[width=0.95\textwidth]{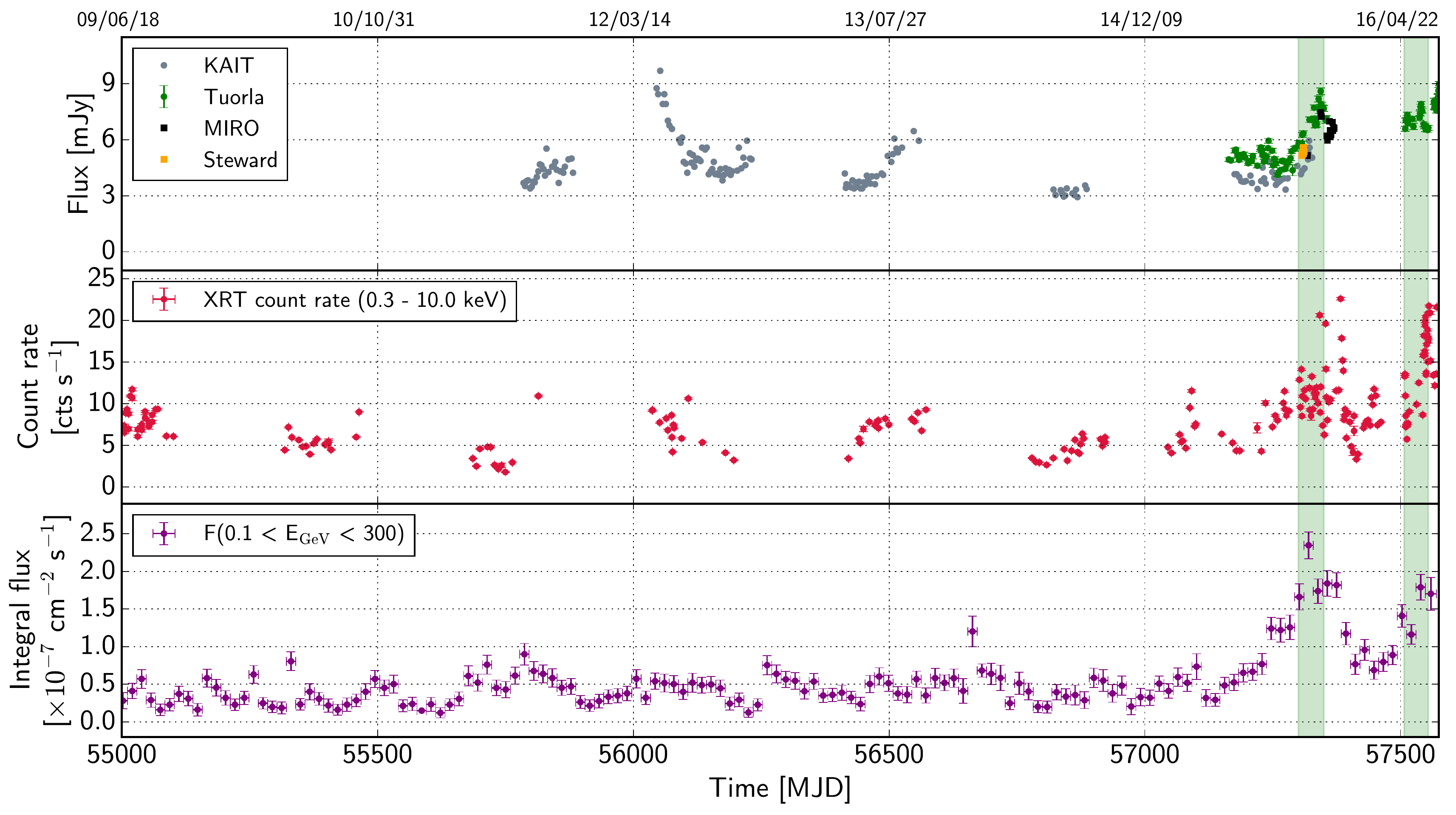}
 \caption{Long-term optical (\emph{top}), \emph{Swift} XRT (\emph{center}) and \emph{Fermi}-LAT (\emph{bottom}) light curves for the blazar 1ES 1959+650. An exceptional period of strong flaring in all bands started in late 2015. The green bands indicate the epochs with VERITAS observations covered in this work.}
  \label{fig:flares}
\end{figure}

In October 2015, the source entered an extended period of flaring activity across multiple wavebands, as can be seen in the historical light curve for optical \cite{TuorlaWeb,Tuorla,KAIT,Steward,MIRO}, \emph{Swift}-XRT\footnote{\href{http://www.swift.psu.edu/monitoring/}{http://www.swift.psu.edu/monitoring/}}~\cite{Swift} and \emph{Fermi}-LAT data shown in Fig.~\ref{fig:flares}. Between October 2015 and June 2016 the blazar was observed with the VERITAS gamma-ray telescope array, and detected in several occasions at a flux $>$1 C.U. We summarize preliminary results from these observations in the following sections.

\section{Detectors and data sets}

Observations of 1ES 1959+650 were conducted using VERITAS, an array of four imaging air Cherenkov telescopes located at the Fred Lawrence Whipple Observatory (FLWO) in southern Arizona (31$^{\circ}$ 40'N, 110$^{\circ}$ 57'W, 1.3km a.s.l.) that is sensitive to gamma rays in the energy band from 85 GeV to $>$30 TeV. The angular resolution of the array (for 68\% containment) is $<$0.1$^{\circ}$ at 1 TeV, and the absolute pointing accuracy of the instrument is better than 50 arcsec. The energy resolution of VERITAS is 15-25\% in $\Delta E/E$. Each VERITAS telescope has a 12-m diameter primary mirror consisting of 345 facets with a Davies-Cotton optical design that concentrates the Cherenkov light produced by air showers onto a camera equipped with 499 photomultiplier tubes (PMTs) covering a $3.5^{\circ}$-diameter field of view. VERITAS can detect the Crab nebula at a significance $>5\sigma$ in about two minutes of observations and a source with gamma-ray flux of 1\% of the Crab in $\sim$ 25 hours.

VERITAS observed 1ES 1959+650 on two epochs: between 2015 October 8 and November 21 UTC (MJD 57303-57347), and between 2016 April 29 and June 16 UTC (MJD 57507-57555). A total exposure of 16.8 hours of quality-selected data was accumulated during the first epoch, including observations using reduced high voltage (RHV) on the PMT camera due to the presence of Moon light, and of 15.0 hours during the second epoch (with no RHV). Operating the cameras in RHV mode effectively lowers the gain of the PMTs, which increases the  energy threshold above which the detector is sensitive to gamma-ray showers, and consequently above which the flux of a source can be claimed. For this reason, new instrument response functions that describe the sensitivity of VERITAS for the 2015 observations were generated and validated to calibrate these observations.

The VERITAS observations were performed using the standard ``wobble'' strategy where the telescopes are offset from the position of the source by $0.5^{\circ}$ to allow for a simultaneous determination of the background. Both data sets show a strong detection of the source with an statistical significance of $82\sigma$ in 2015  and $80\sigma$ in 2015. During this period VERITAS circulated three Astronomer's Telegrams to the community together with other instruments (\#8148\footnote{\url{http://www.astronomerstelegram.org/?read=8148}} on 2015 October 15, \#9010\footnote{\url{http://www.astronomerstelegram.org/?read=9010}} on 2016 April 30, and \#9148\footnote{\url{http://www.astronomerstelegram.org/?read=9148}} on 2016 June 13) to encourage multiwavelength follow-up observations of these rare, strong flaring events. 

\section{Analysis results and discussion}

The analysis of VERITAS data involves introducing cuts to separate gamma-ray shower candidate events from a dominant background of hadronic cosmic-ray showers. These cuts are applied on the parameters that characterize the geometry of the shower images. The background contribution of the events passing these cuts is estimated using the ring background method. Sky maps are constructed by correlating event positions using a circular top-hat function with the radius of the VERITAS point-spread function. The statistical significance of deviations with respect to the background is calculated using the method of Li\&Ma. We here present light curves and energy spectra of 1ES 1959+650 during the two observational epochs in 2015 and 2016. 

\subsection{Light curves}

\begin{figure}[th]
\centering
\includegraphics[width=0.75\textwidth]{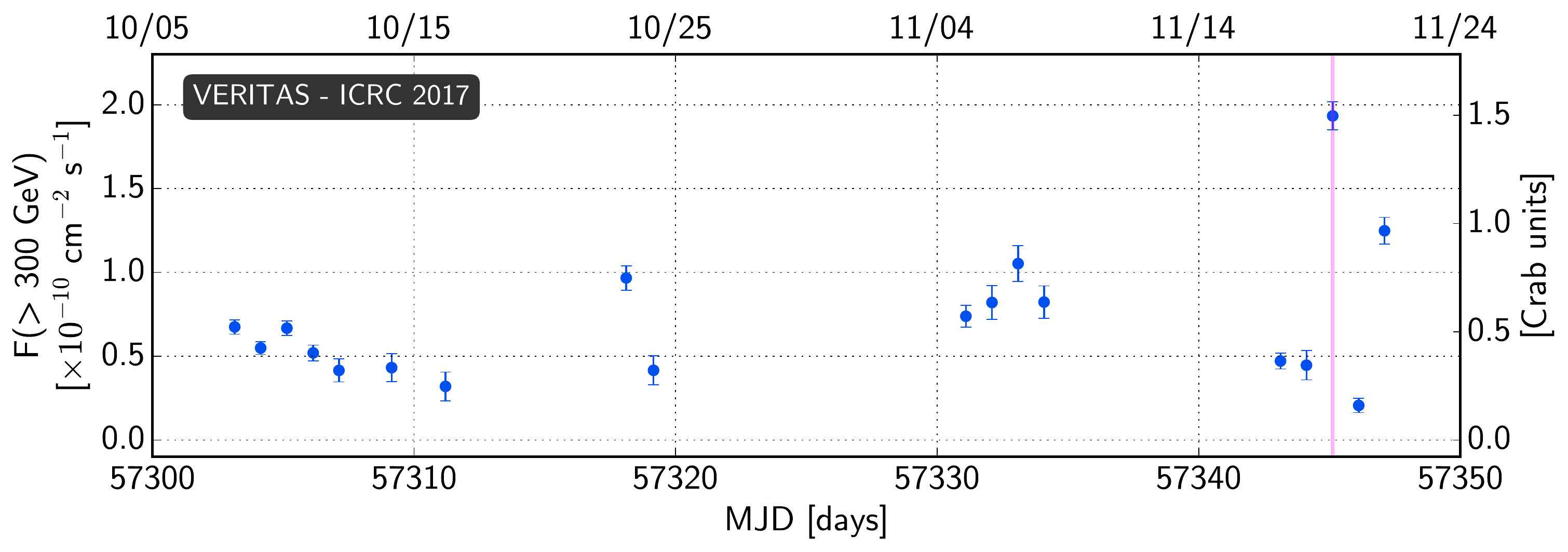}
\includegraphics[width=0.75\textwidth]{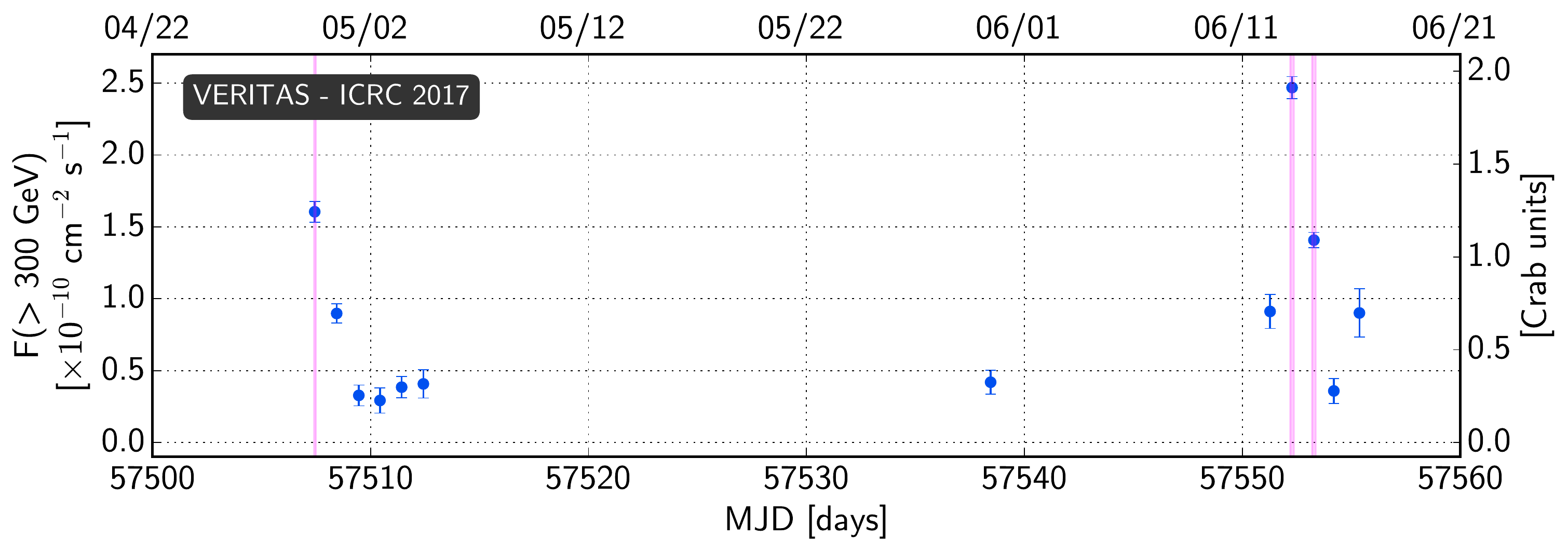}
 \caption{Nightly integral flux light curves for 2015 (\emph{top}) and 2016 (\emph{bottom}) above an energy threshold of 300 GeV. The axis on the right shows the flux in Crab units above the same energy threshold. The four periods with an average flux higher than 1 C.U. are highlighted in magenta and detailed light curves for those periods are shown in the next figure. }
  \label{fig_daily_lc}
\end{figure}

Nightly integral flux light curves were computed above an energy threshold of 300 GeV for the 2015 and 2016 epochs, which are shown in Fig.~\ref{fig_daily_lc}. 

\begin{figure}[th]
\centering
\includegraphics[width=0.45\textwidth]{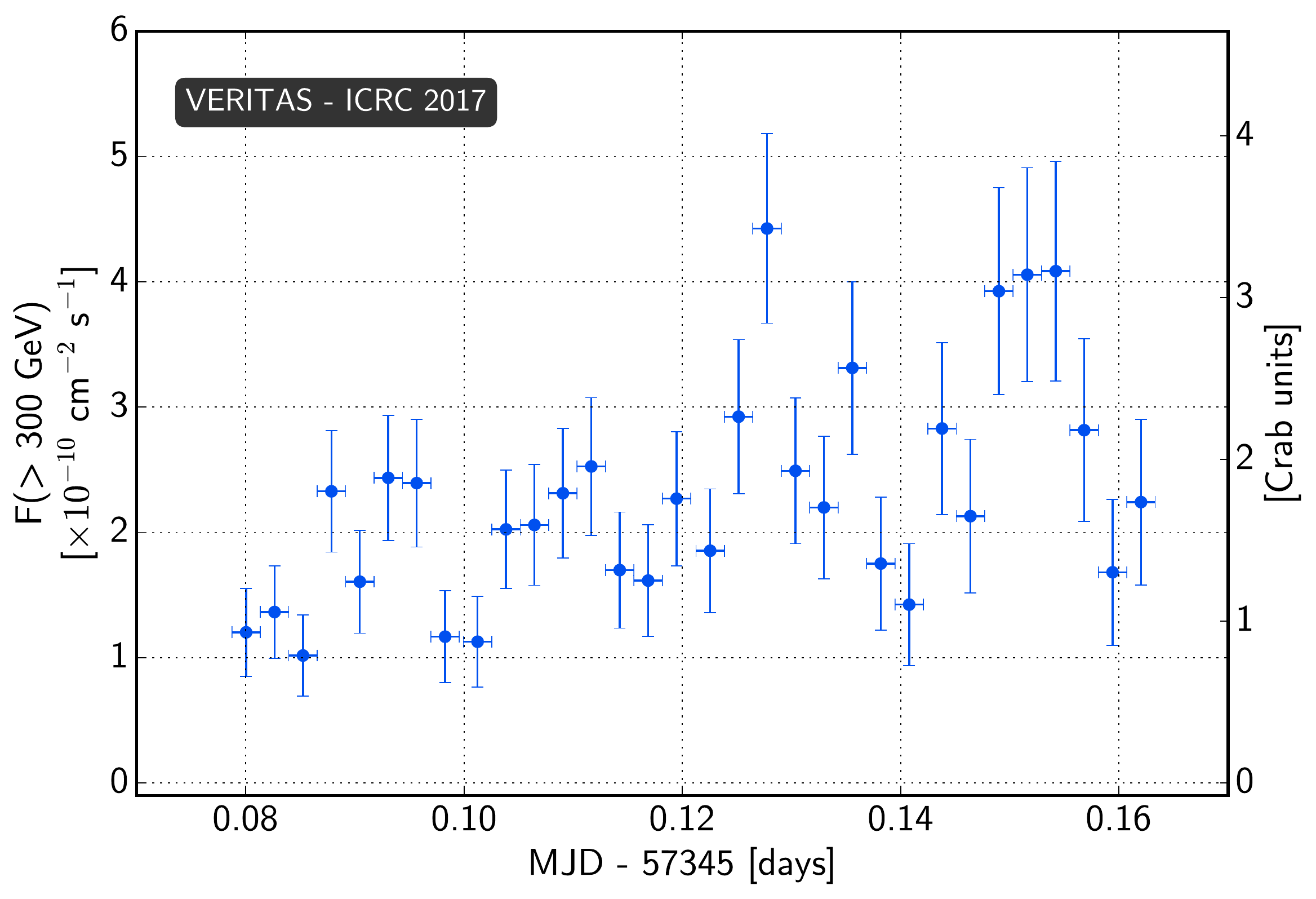}
\includegraphics[width=0.45\textwidth]{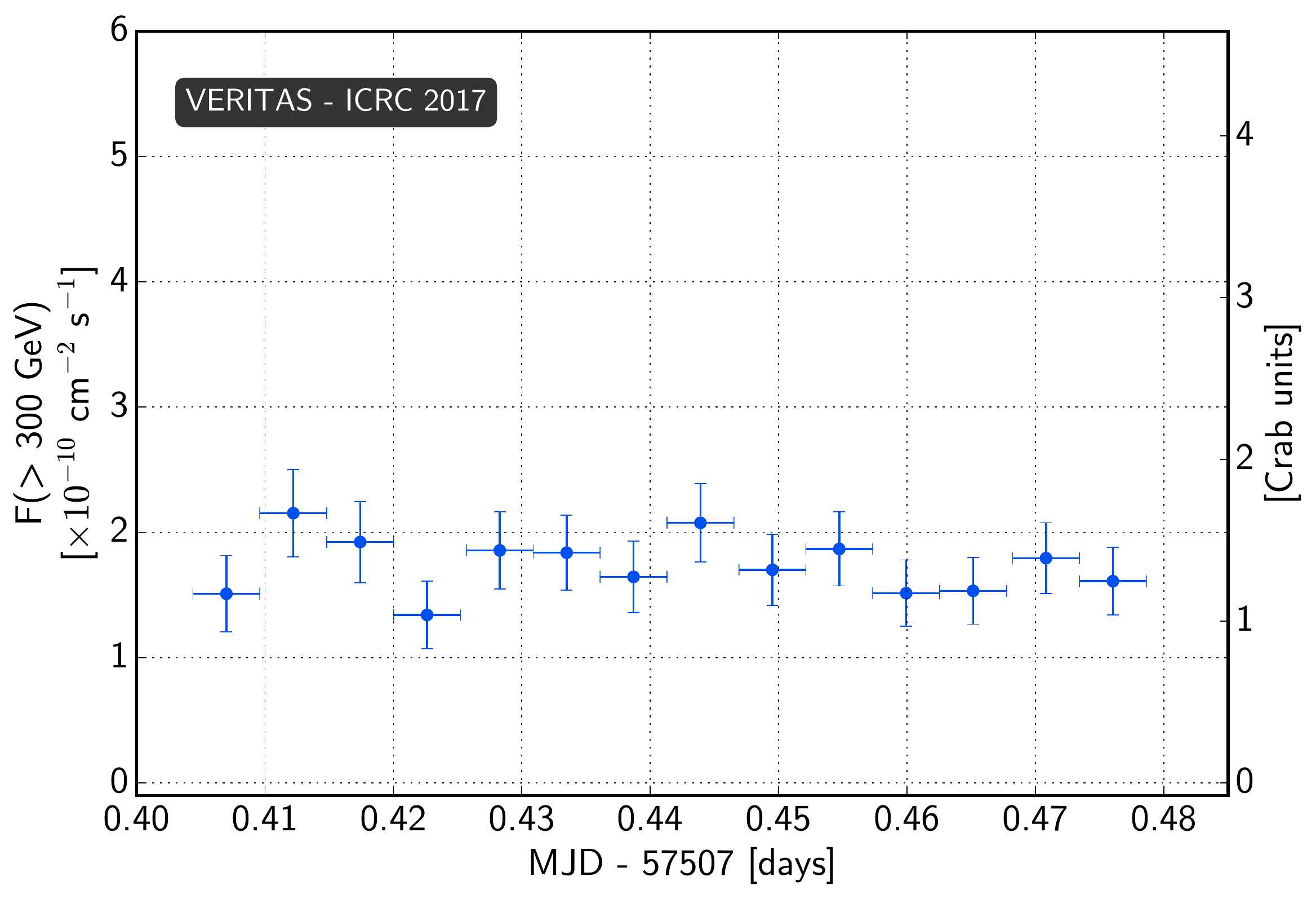}
\includegraphics[width=0.45\textwidth]{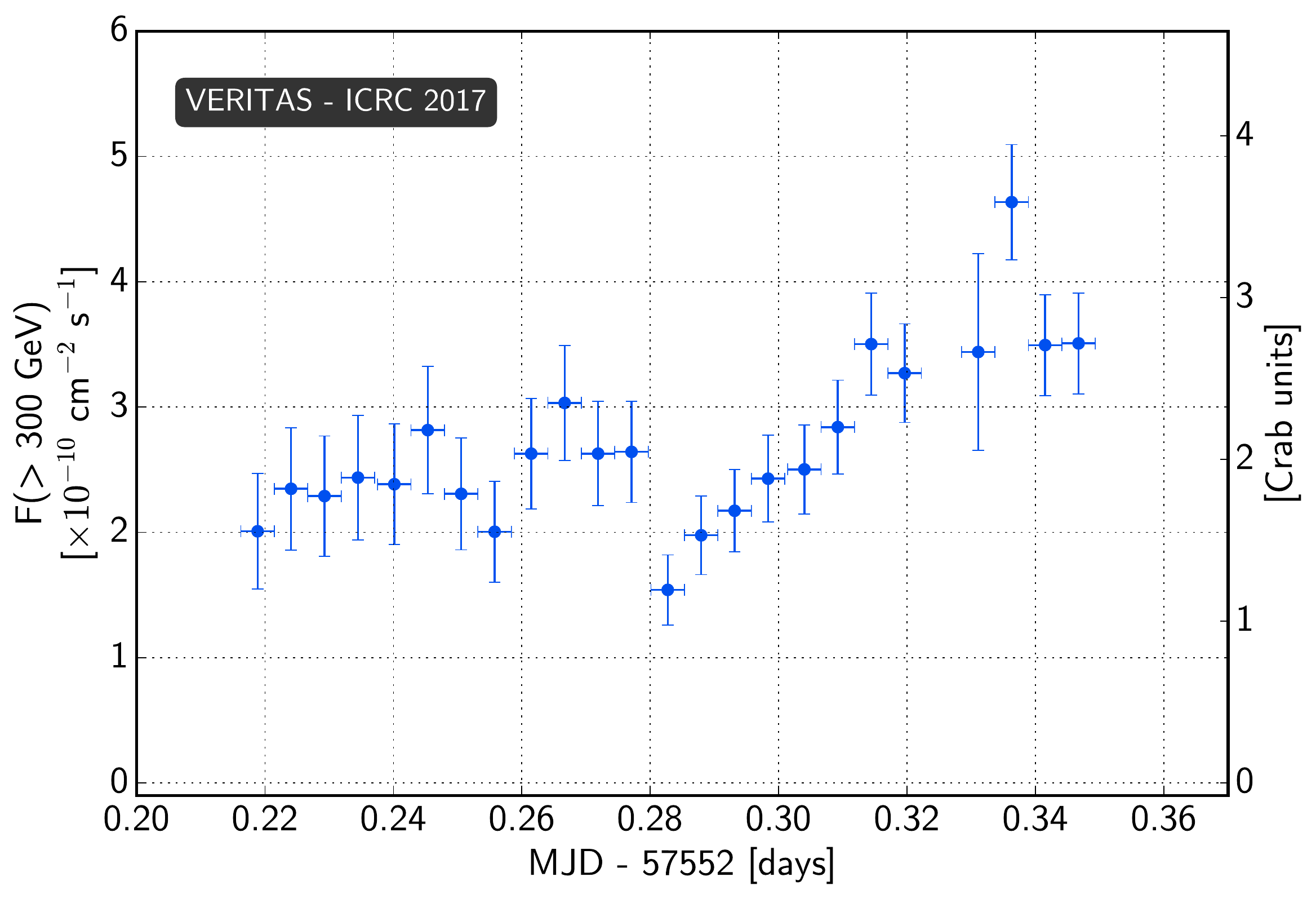}
\includegraphics[width=0.45\textwidth]{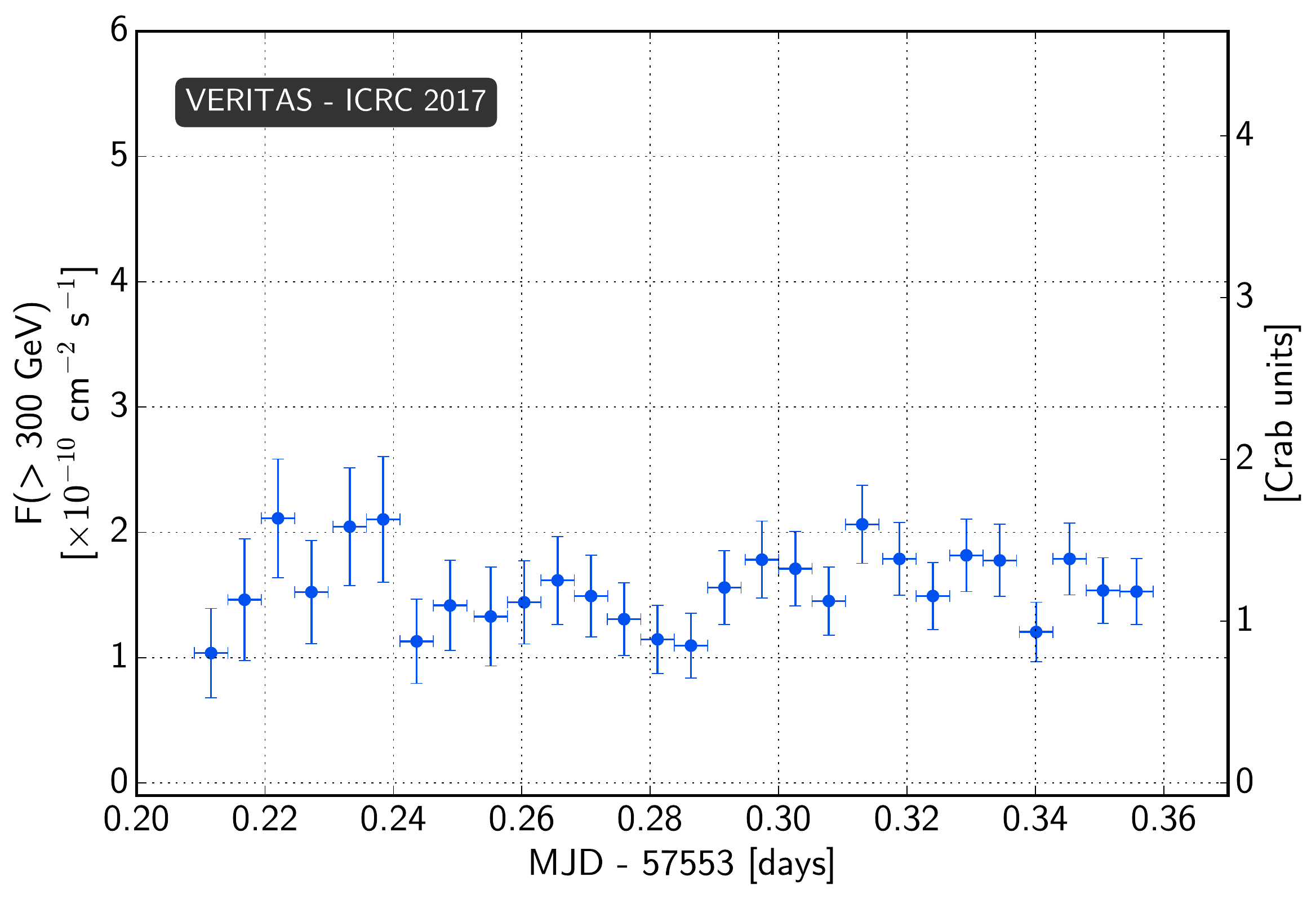}
 \caption{Detailed light curves with 8-minute time bins for the nights where the average flux was above 1 C.U.: MJD 57345 (\emph{top left}), MJD 57507 (\emph{top right}), MJD 57552 (\emph{bottom left}), and MJD 57553 (\emph{bottom right}). Intra-night variability is evident on MJD 57345 and 57552.}
  \label{fig_run_lc}
\end{figure}

During the top epochs, the average nightly flux was higher than 1 C.U. (or equivalently a flux of $1.29 \times 10^{-10}$ photons TeV$^{-1}$ cm$^{-2}$ s$^{-1}$) on four separate nights of VERITAS observations: 2015 November 19 UTC (MJD 57345), 2016 April 29 UTC (MJD 57507), 2016 June 13 UTC (MJD 57552), and 2016 June 14 UTC (MJD 57553). We present in Fig.~\ref{fig_run_lc} detailed light curves with 8-minute time bins for those nights.

\subsection{Energy spectra}

\begin{figure}[th]
\centering
\includegraphics[width=0.45\textwidth]{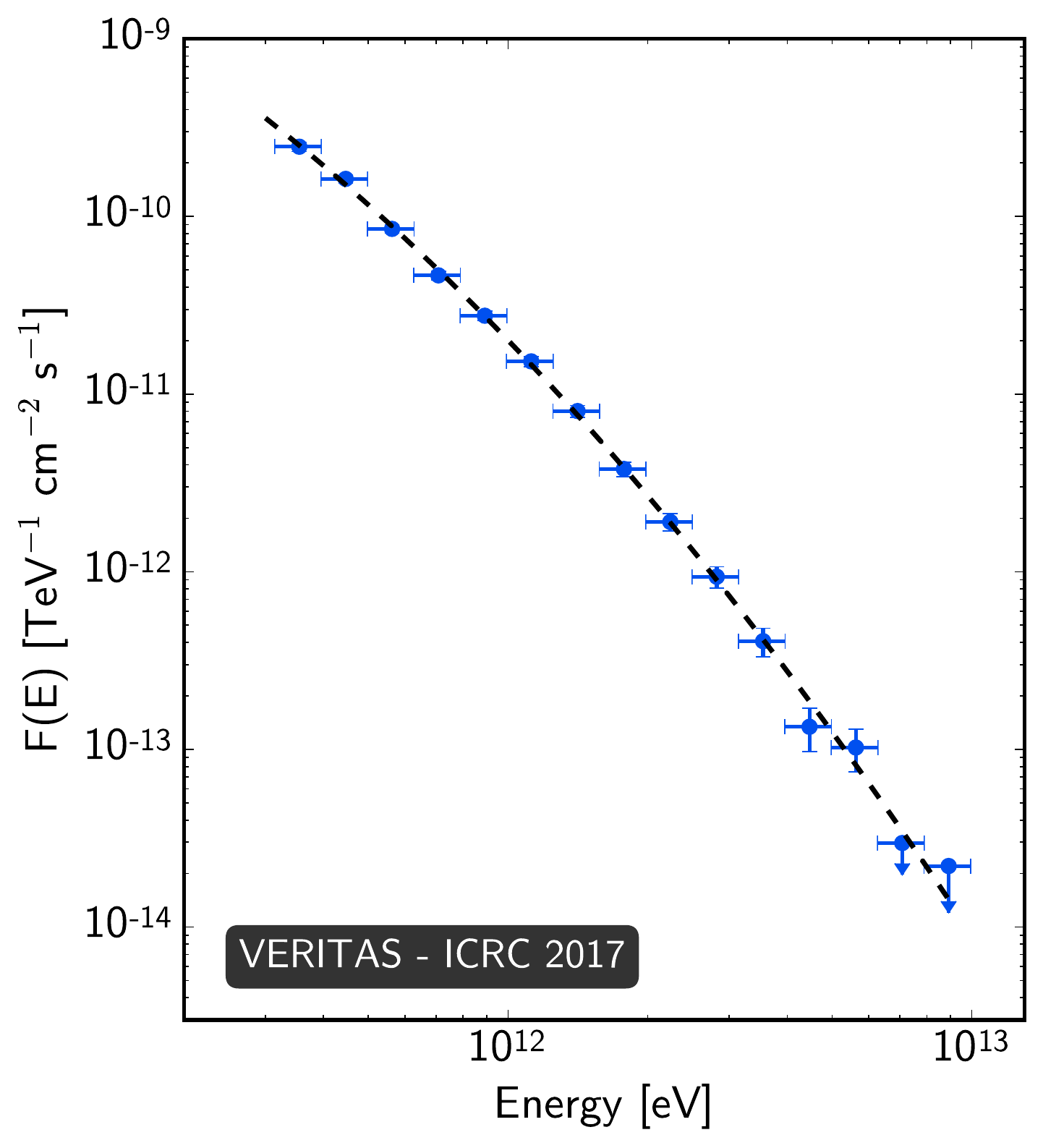}
\includegraphics[width=0.45\textwidth]{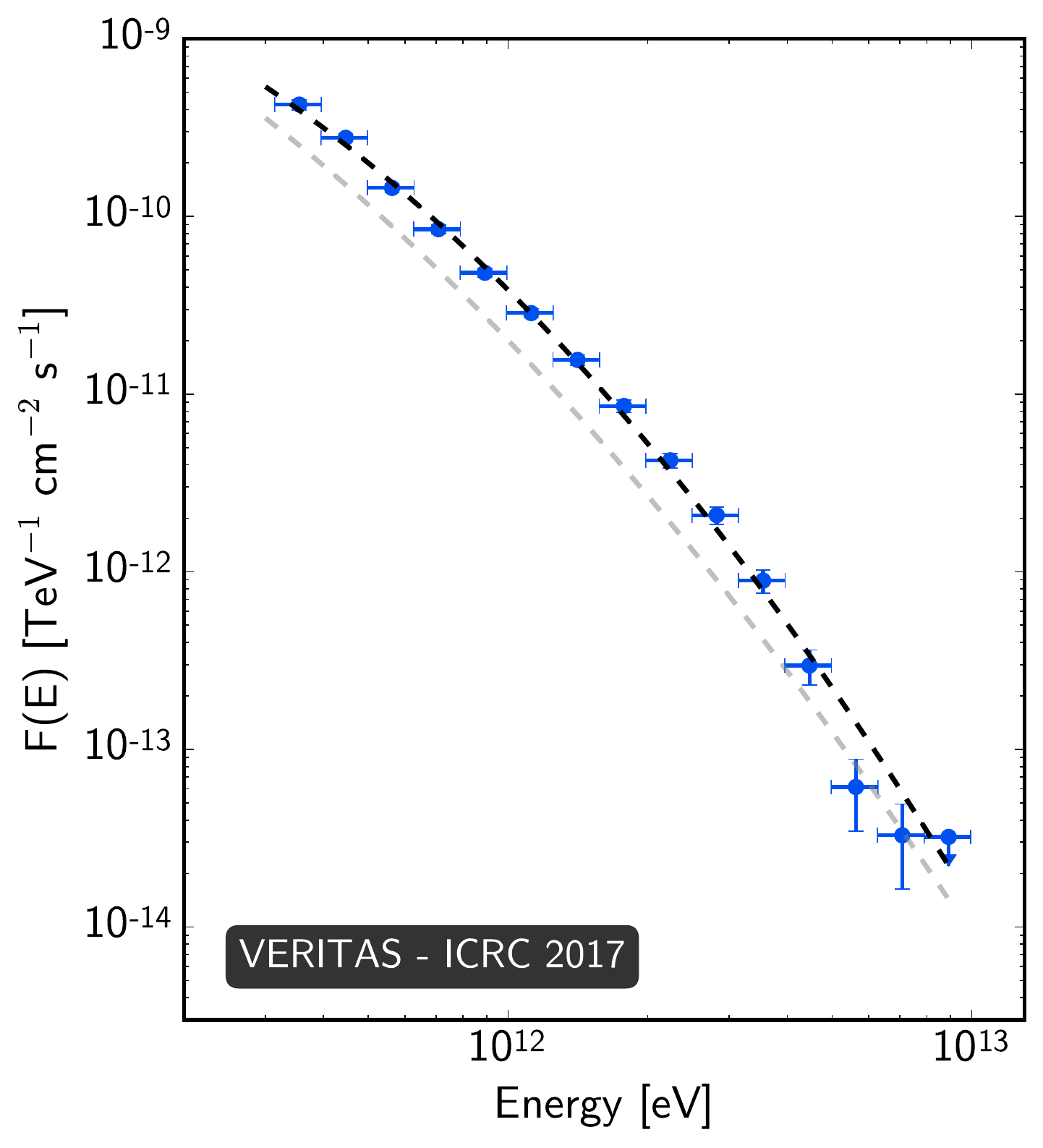}
 \caption{Differential energy spectra for 1ES 1959+650 as measured by VERITAS during the 2015 (\emph{left}) and 2016 (\emph{right}) observation periods. The spectra are fit with log-parabolic functions, shown as dashed black lines. For comparison, the 2015 best-fit spectrum is shown in gray in the 2016 spectrum.}
  \label{fig_spectra}
\end{figure}

Time-averaged differential spectra above 300 GeV of 1ES 1959+650 are shown in Fig.~\ref{fig_spectra} for the two observational epochs with ten flux bins per energy decade. A power-law fit was performed in the energy range between 300 GeV and 10 TeV using a function of the form $F(E) = A (E/E_{0})^{-\alpha}$ where $A$ is the flux at the normalization energy $E_{0}$ of 1 TeV and $\alpha$ is the index. 

A bad fit was obtained for this simple power-law model, as evidenced by large reduced-$\chi^2$ values, due to the clear curvature of both spectra visible in Fig.~\ref{fig_spectra}. A fit was performed using a log-parabolic function of the form $F(E) = A (E/E_{0})^{-\alpha - \beta \log(E/E_{0})}$, where $\beta$ represents the curvature. Table~\ref{table_fit} summarizes the best-fit parameters and corresponding reduced $\chi^2$ for the two functional models. The 2015 spectrum is well-fit by the log-parabolic function (fit $p$-value: 70\%). The 2016 log-parabolic fit, although strongly favored over the power-law fit, still shows some tension with the data ($p$-value: 0.6\%). This could represent evidence of spectral variability during the epoch, which will be studied in detail in the future.

\begin{table}[t]
\centering
\begin{tabular}{cccccc}
\toprule
Year & Fit function & $A$ & $\alpha$ & $\beta$ &  $\chi^2$/ndf  \\
& & [$\times 10^{-11}$ TeV$^{-1}$ cm$^{-2}$ s$^{-1}$] & & &  \\
\hline
2015 & PL & $1.74 \pm 0.04$ & $2.75 \pm 0.03$ & ---  & 49.0/11 \\
2015 & LP & $2.01 \pm 0.06$ & $2.72 \pm 0.03$ & $0.63 \pm 0.11$ & 7.3/10 \\
\hline
2016 & PL & $3.05 \pm 0.06$ & $2.79 \pm 0.02$ & --- &  147.9/12 \\
2016 & LP & $3.86 \pm 0.10$ & $2.62 \pm 0.03$ & $0.83 \pm 0.09$ & 26.4/11 \\
\bottomrule
\end{tabular}
\caption{Best-fit parameters for the observational epochs of 2015 and 2016 using simple power law (PL) or log-parabola (LP) functions. The improvement in reduced $\chi^{2}$ for the LP model is evident.}
\label{table_fit}
\end{table}

\section{Conclusions}

We have presented a summary of the 2015 and 2016 VERITAS observations of strong flaring events from the blazar 1ES 1959+650. The source was observed on multiple occasions at a flux level higher than the Crab nebula above an energy threshold of 300 GeV reaching a flux level of $\sim3$ C.U. and displaying intra-night variability.
The energy spectrum of the source has a spectral index of $\sim 2.7$, in good agreement with previous observations, and shows a curvature that can be well characterized by a log-parabolic function.

A detailed analysis of the spectral evolution of the source during these flares as well as an analysis of multi-wavelength data will be presented in a  publication currently in preparation.

\section{Acknowledgements}
This work makes use of optical photometry data from the Tuorla blazar monitoring program~\cite{TuorlaWeb, Tuorla}, the KAIT Fermi AGN Light-Curve Reservoir~\cite{KAIT}. Data from the Steward Observatory spectropolarimetric monitoring project were used~\cite{Steward}, which is supported by Fermi Guest Investigator grants NNX08AW56G, NNX09AU10G, NNX12AO93G, and NNX15AU81G.

VERITAS is supported by grants from the U.S. Department of Energy Office of Science, the U.S. National Science Foundation and the Smithsonian Institution, and by NSERC in Canada. We acknowledge the excellent work of the technical support staff at the Fred Lawrence Whipple Observatory and at the collaborating institutions in the construction and operation of the instrument. The VERITAS Collaboration is grateful to Trevor Weekes for his seminal contributions and leadership in the field of VHE gamma-ray astrophysics, which made this study possible.

\end{document}